\documentclass{article}
\usepackage{bm}
\begin{document}
\title{Nonzero $\theta_{13}$ and CP violation from broken $\mu-\tau$ symmetry\\ with $m_{1}=0$\footnote{Talk given at XXX$^{th}$ International Workshop on High Energy Physics "Particle and Astroparticle Physics, Gravitation and Cosmology: Predictions, Observations and New Projects", Protvino, Russia, 23-27 June 2014}}
\author{\bf{Asan Damanik}\footnote{E-mal: d.asan@lycos.com}\\Faculty of Science and Technology\\Sanata Dharma University\\Kampus III USD Paingan Maguwoharjo Sleman Yogyakarta\\Indonesia}
\date{}

\maketitle

\begin{abstract}
Nonzero of mixing angle $\theta_{13}$ has some phenomenological consequences on neutrino physics beyond the standard model.  If the mixing angle $\theta_{13}\neq 0$, then there is the possibility of the CP violation existence on the neutrino sector.  To obtain a nonzero of mixing angle $\theta_{13}$ from neutrino mass matrix obey $\mu-\tau$, we break it by introducing one small parameter $x$ into neutrino mass matrix and then calculated  the Jarlskog invariant as a measure of CP violation existence using the reported experimental data as input and put $m_{1}=0$ for neutrino mass in normal hierarchy.
\end{abstract}

\section{Introduction}
Since the first experimental detected the effect of neutrino oscillation (deficit in the flux of solar neutrino with respect to the Standard Model Solar prediction) in Davis's Homestake Experiment in the late 1960s up today, we can see that the concept of neutrino and our understanding on neutrino should be changed and we must go beyond the standard model of particle physics.    There are some major long standing problems in neutrino physics i.e. neutrino masses (absolute value, hierarchy, mechanism of mass generation), the underlying symmetry of neutrino mass matrix, and the question related to the kind of neutrino from the particle-antiparticle aspect (whether it Majorana or Dirac particle).

Related to the neutrino mass matrix and its underlying symmetry, recently, there are some symmetries proposed by many authors i.e. $U(1)_{L'}$ based on conservation of $L_{e}-L_{\mu}-L_{\tau}$ \cite{Petcov, Leung, Zee, Barbieri} and $\mu-\tau$ symmetry based on the invariance of the flavor neutrino mass terms under the interchange of $\nu_{\mu}$ and $\nu_{\tau}$ \cite{Fukuyama, Koide, Koide1, Matsuda, Matsuda1, Matsuda2, Matsuda3, Mohapatra3, Lam, Lam1, Ma, Datta, Kitabayashi, Kitabayashi1, Kitabayashi2, Aizawa1, Aizawa2, Grimus, Grimus1, Grimus2, Grimus3, Mohapatra4, Mohapatra5}.  Among the proposed underlying symmetries of neutrino mass matrix, the $\mu-\tau$ symmetry is the most intereting symmetry because the $\mu-\tau$ symmetry reduces the number of parameters in symmetric neutrino mass matrix from 6 parameters to 4 parameters and it can also be derived from the three well-known neutrino mixing matrices i.e. tribimaximal (TBM), bimaximal (BM), dan democratic (DC) with some approximations.  But, the three well-known of neutrino mixing matrices predict mixing angle $\theta_{13}=0$ which is not in agreement with the recent experimental results which indicate the mixing angle $\theta_{13}\neq 0$. 

The current combined world data for neutrino squared-mass difference are given by \cite{Gonzales-Carcia, Fogli}:
\begin{equation}
\Delta m_{21}^{2}=7.59\pm0.20 (_{-0.69}^{+0.61}) \times 10^{-5}~\rm{eV^{2}},\label{M21}
\end{equation}
\begin{equation}
\Delta m_{32}^{2}=2.46\pm0.12(\pm0.37) \times 10^{-3}~\rm{eV^{2}},~\rm(for~ NH)\label{m32}
\end{equation}
\begin{equation}
\Delta m_{32}^{2}=-2.36\pm0.11(\pm0.37) \times 10^{-3}~\rm{eV^{2}},~\rm(for~ IH)\label{m321}
\end{equation}
\begin{equation}
\theta_{12}=34.5\pm1.0 (_{-2.8}^{3.2})^{o},~~\theta_{23}=42.8_{-2.9}^{+4.5}(_{-7.3}^{+10.7})^{o},~~\theta_{13}=5.1_{-3.3}^{+3.0}(\leq 12.0)^{o},
 \label{GD1}
\end{equation}
at $1\sigma~(3\sigma)$ level.

In order to accommodate the nonzero $\theta_{13}$ and the Jarlskog rephasing invariant $J_{\rm CP}$ as a measure of CP violation in neutrino sector, in this talk we break the $\mu-\tau$ symmetry by introducing one small parameter to break the neutrino mass matrix that obey $\mu-\tau$ symmetry.  The paper is organized as follow: in section 2 we break the neutrino mass matrix with $\mu-\tau$ symmetry by introducing a small parameter into neutrino mass matrix and in section 3 we determine the $J_{\rm CP}$ by using the experimental results as input.  Finally, section 4 is devoted to conclusions.

\section{Broken $\mu-\tau$ symmetry with trace remain constant}
In the basis where the charged lepton mass matrix is diagonal, the neutrino mass matrix $M_{\nu}$ can be diagonalized by mixing matrix $V$ as follow:
\begin{eqnarray}
M_{\nu}=VMV^{T},\label{m}
\end{eqnarray}
where the diagonal neutrino mass matrix $M=Diag (m_{1},m_{2},m_{3})$ and the mixing matrix $V$ in the standard parametrization is given by:
\begin{equation}
V=\bordermatrix{& & &\cr
&c_{12}c_{13} &s_{12}c_{13} &s_{13}e^{-i\delta}\cr
&-s_{12}c_{23}-c_{12}s_{23}s_{13}e^{i\delta} &c_{12}c_{23}-s_{12} s_{23}s_{13}e^{i\delta}&s_{23}c_{13}\cr
&s_{12}s_{23}-c_{12}c_{23}s_{13}e^{i\delta} &-c_{12}s_{23}-s_{12}c_{23}s_{13}e^{i\delta} &c_{23}c_{13}}
 \label{Vv}
\end{equation}
where $c_{ij}$ is the $\cos\theta_{ij}$, $s_{ij}$ is the $\sin\theta_{ij}$, and $\theta_{ij}$ are the mixing angles. 
If we put the mixing angle $\theta_{13}=0$, consequently $s_{13}=0$ and $c_{13}=1$ in Eq. (\ref{Vv}), then the neutrino mass matrix $M_{\nu}$ in Eq. (\ref{m}) read:
\begin{eqnarray}
M_{\nu}=\bordermatrix{& & &\cr
&P &Q &-Q\cr
&Q &R &S\cr
&-Q &S &T},\label{M}
\end{eqnarray}
where:
\begin{eqnarray}
P=m_{1}c_{12}^{2}+m_{2}s_{12}^{2},\label{P1}\\
Q=(m_{2}-m_{1})c_{12}s_{12}c_{23},\label{Q1}\\
R=(m_{1}s_{12}^{2}c_{23}^{2}+m_{2}c_{12}^{2})c_{23}^{2}+m_{3}s_{23}^{2},\label{R1}\\
S=(-m_{1}s_{12}^{2}-m_{2}c_{12}^{2}+m_{3})s_{23}c_{23}.\label{S1}\\
T=m_{1}s_{12}^{2}s_{23}^{2}+m_{2}c_{12}^{2}s_{23}^{2}+m_{3}c_{23}^{2}. \label{T1}
\end{eqnarray}

One can see in Eq. (\ref{M}) that the neutrino mass matrix ($M_{\nu}$) deduced from neutrino mixing matrix with assumption $\theta_{13}=0$ and $\theta_{23}=\pi/2$ give: $S=T$ and the resulted neutrino mass matrix is the $\mu-\tau$ symmetry.  But, as dictated from the experimental results, the mixing angles: $\theta_{23}\neq\pi/2$ and  $\theta_{13}\neq 0$ and relatively large \cite{Double, Minos, T2K, Daya, RENO} which imply that the assumption: $\theta_{23}=\pi/2$ and $\theta_{13}=0$ in formulating the neutrino mixing matrix must be rule out and hence the exact $\mu-\tau$ symmetry as the underlying symmetry of neutrino mass matrix is no longer adequate to accommodate the recent experimental results.

Concerning the neutrino mass matrix that obey $\mu-\tau$  symmetry and mixing angle $\theta_{13}$, Mohapatra \cite{Mohapatra4} stated explicitly that neutrino mass matrix which obey $\mu-\tau$ symmetry to be the reason for maximal $\mu-\tau$ mixing and one gets $\theta_{13}=0$, conversely if $\theta_{13}\neq 0$ can provide the $\mu-\tau$ symmetry beraking manifests in the case of normal hierarchy.  Aizawa and Yasue \cite{Aizawa} analysis complex neutrino mass texture and the $\mu-\tau$ symmetry which can yield small $\theta_{13}$ as a $\mu-\tau$ breaking effect.   The $\mu-\tau$ symmetry breaking effect in relation with the small $\theta_{13}$ also discussed in \cite{Fuki}.  Analysis of the correlation between CP violation and the $\mu-\tau$ symmetry breaking can be read in \cite{Mohapatra2, Baba, He1, Damanik}.

Now, we are in position to study the effect of neutrino mass matrix that obey the $\mu-\tau$ symmetry breaking in relation to the Jarlskog rephasing invariant $J_{\rm CP}$ by breaking the neutrino mass matrix in Eq. (\ref{M}).  We break the neutrino mass matrix in Eq. (\ref{M}) by introducing a small parameter $x$ with the constraint that the trace of the broken neutrino mass matrix is remain constant with the unbroken one.  This scenario of breaking has been applied in Ref. \cite{Damanik1} to break the neutrino mass matrix invariant under a cyclic permutation.  In this breaking scenario, the broken neutrino mass matrix obtained from neutrino mass matrix obey $\mu-\tau$ symmetry reads \cite{Damanik4}:
\begin{equation}
M_{\nu}=\bordermatrix{& & &\cr
&P &Q &Q\cr
&Q &R-ix &S\cr
&Q &S &R+ix}.\label{M1}
\end{equation}

As stated previously that the CP violation can be determined from the Jarlskog rephasing invariant $J_{\rm CP}$.  Alternatively, Jarlskog rephasing invariant $J_{\rm CP}$ can be determined using the relation \cite{Branco}:
\begin{equation}
J_{\rm CP}=-\frac{{\rm Im}\left[(M_{\nu}^{'})_{e\mu}(M_{\nu}^{'})_{\mu\tau}(M_{\nu}^{'})_{\tau e}\right]}{\Delta m_{21}^{2}\Delta m_{32}^{2}\Delta m_{31}^{2}}, \label{JCP}
\end{equation}
where $(M_{\nu}^{'})_{ij}=(M_{\nu}M_{\nu}^{\dagger})_{ij}$ with $ i,j=e,\nu,\tau$, and $\Delta m_{kl}^{2}=m_{k}^{2}-m_{l}^{2}$ with $k=2,3,~ {\rm and}~ l=1,2$.  From Eq. (\ref{M1}), we have:
\begin{equation}
M_{\nu}^{'}=\bordermatrix{& & &\cr
&P^{2}+2Q^{2} &Q(P+S+R+ix) &Q(P+R+S-ix)\cr
&Q(P+S+R-ix) &Q^{2}+R^{2}+S^{2}+x^{2} &Q^{2}+2S(R-ix)\cr
&Q(P+S+R+ix) &Q^{2}+2S(R+ix) &Q^{2}+R^{2}+S^{2}+x^{2}}. \label{M2}
\end{equation}

\section{Nonzero $\theta_{13}$ and Jarlskog rephasing invariant}
From Eqs. (\ref{JCP}) and (\ref{M2}) we have the Jarlskog rephasing invariant as follow:
\begin{equation}
J_{\rm CP}=\frac{2Q^{2}\left[S(S+P)^{2}-Q^{2}(R+S+P)-R^{2}S\right]x-(2Q^{2}S)x^{3}}{\Delta m_{21}^{2}\Delta m_{32}^{2}\Delta m_{31}^{2}}. \label{JCP1}
\end{equation}
If we insert Eqs. (\ref{P1})-(\ref{S1}) into Eq. (\ref{JCP1}), then we have the $J_{\rm CP}$ as follow:
\begin{eqnarray}
J_{\rm CP}=\frac{2(m_{2}-m_{1})^{2}c_{12}^{2}s_{12}^{2}c_{23}^{2}}{\Delta m_{21}^{2}\Delta m_{32}^{2}\Delta m_{31}^{2}}[[(-m_{1}s_{12}^{2}-m_{2}c_{12}^{2}+m_{3})s_{23}c_{23}\nonumber \\  \times((-m_{1}s_{12}^{2}-m_{2}c_{12}^{2}+m_{3})s_{23}c_{23}+m_{1}c_{12}^{2}+m_{2}s_{12}^{2})^{2}\nonumber \\-(m_{2}-m_{1})^{2}c_{12}^{2}s_{12}^{2}c_{23}^{2}((m_{1}s_{12}^{2}+m_{2}c_{12}^{2})c_{23}^{2}+m_{2}s_{23}^{2}\nonumber \\-(m_{1}s_{12}^{2}+m_{2}c_{12}^{2}-m_{3})s_{23}c_{23}+m_{1}c_{12}^{2}+m_{2}s_{12}^{2})\label{JCP2} \\-((m_{1}s_{12}^{2}+m_{2}c_{12}^{2})c_{23}^{2}+m_{2}s_{23}^{2})^{2}(-m_{1}s_{12}^{2}-m_{2}c_{12}^{2}\nonumber \\+m_{3})s_{23}c_{23}]x-c_{23}s_{23}[-m_{1}s_{12}^{2}-m_{2}c_{12}^{2}+m_{3}]x^{3}].\nonumber
\end{eqnarray}
From Eq. (\ref{JCP2}), one can see that the exact $\mu-\tau$ symmetry ($x=0$) lead to $J_{\rm CP}=0$ as claimed by many authors, and if we still want to have $J_{\rm CP}\neq 0$ from the $\mu-\tau$ symmetry, then we must break it softly.

It is apparent from Eq. (\ref{JCP2}) that in this breaking scenario the Jarlskog rephasing invariant ($J_{\rm CP}$) does not depend on the mixing angle $\theta_{13}$.  Because $x$ is very small, the last term in Eq. (\ref{JCP2}) is very small compare to the first term and then it can be neglected.  By neglecting the last term contribution of Eq. (\ref{JCP2}) to the Jarlskog rephasing invariant, we have:
\begin{eqnarray}
J_{\rm CP}\approx\frac{2(m_{2}-m_{1})^{2}c_{12}^{2}s_{12}^{2}c_{23}^{2}}{\Delta m_{21}^{2}\Delta m_{32}^{2}\Delta m_{31}^{2}}[(-m_{1}s_{12}^{2}-m_{2}c_{12}^{2}+m_{3})s_{23}c_{23}\nonumber \\  \times((-m_{1}s_{12}^{2}-m_{2}c_{12}^{2}+m_{3})s_{23}c_{23}+m_{1}c_{12}^{2}+m_{2}s_{12}^{2})^{2}\nonumber \\-(m_{2}-m_{1})^{2}c_{12}^{2}s_{12}^{2}c_{23}^{2}((m_{1}s_{12}^{2}+m_{2}c_{12}^{2})c_{23}^{2}+m_{2}s_{23}^{2}\nonumber \\-(m_{1}s_{12}^{2}+m_{2}c_{12}^{2}-m_{3})s_{23}c_{23}+m_{1}c_{12}^{2}+m_{2}s_{12}^{2}) \label{JCP3}\\-((m_{1}s_{12}^{2}+m_{2}c_{12}^{2})c_{23}^{2}+m_{2}s_{23}^{2})^{2}(-m_{1}s_{12}^{2}-m_{2}c_{12}^{2}\nonumber \\+m_{3})s_{23}c_{23}]x.\nonumber
\end{eqnarray}

In order to get the value of Jarlskog rephasing invariant $J_{\rm CP}$ of Eq. (\ref{JCP3}), we use the experimental values of neutrino oscillation in Eqs. (\ref{M21})-(\ref{GD1}) as input.  We also put $m_{1}=0$ as a first approximation and within this scenario, for normal hierarchy (NH):
\begin{eqnarray}
m_{2}^{2}=\Delta m_{21}^{2},\\
m_{3}^{2}=\Delta m_{32}^{2}+\Delta m_{21}^{2}.\label{m3}
\end{eqnarray}

By using the the experimental values of neutrino as shown in Eqs. (\ref{M21})-(\ref{GD1}) as input into Eq. (\ref{JCP3}), then we have:
\begin{eqnarray}
J_{\rm CP}\approx 0.4644x.\label{JCP4}
\end{eqnarray}
If we determine the Jarlskog rephasing invariant from neutrino mixing matrix of Eq. (\ref{Vv}), by using the relation:
\begin{eqnarray}
J_{\rm CP}={\rm Im} (V_{11}^{*}V_{23}^{*}V_{13}V_{21}),
\end{eqnarray}
then we have:
\begin{eqnarray}
J_{\rm CP}=c_{12}s_{12}c_{23}s_{23}c_{13}^{2}s_{13}\sin{\delta}.\label{AD1}
\end{eqnarray}
As indicated by the experimental fact that the mixing angle $\theta_{13}$ is very small, we can approximate $c_{13}\approx 1$ and Eq. (\ref{AD1}) read:
\begin{eqnarray}
J_{\rm CP}\approx c_{12}s_{12}c_{23}s_{23}s_{13}\sin{\delta}.\label{AD2}
\end{eqnarray}

By inserting the experimental values of mixing angles in Eq. (\ref{GD1}) (nonzero $\theta_{13}$ and $\theta_{23}\neq\pi/2$) into Eq. (\ref{AD2}) and equate it with Eq. (\ref{JCP4}), we have the Dirac phase ($\delta$) depend on perturbed parameter $x$:
\begin{eqnarray}
\sin\delta\approx\frac{0.4644x}{0.0207},
\end{eqnarray}
or
\begin{eqnarray}
x\approx 0.0446\sin\delta.
\end{eqnarray}

\section{Conclusions}
We have studied systematically the breaking effect on neutrino mass that obey $\mu-\tau$ symmetry by introducing a small parameter $x$ into neutrino mass matrix with the requirement that the trace of the broken $\mu-\tau$ symmetry is remain constant.  By using the experimental data of neutrino oscillations as input and put $m_{1}=0$, we can obtain the Jarlskog rephasing invariant $J_{\rm CP}\neq 0$ which indicate the existence of CP violation in neutrino sector and the Dirac phase $\delta$ depend on the parameter $x$ for neutrino in normal hierarchy.

\section*{Acknowledgment}
Author would like to thank to the Organizer of the XXX$^{th}$ International Workshop on High Energy Physics "Particle and Astroparticle Physics, Gravitation and Cosmology: Predictions, Observations and New Projects" for a nice hospitality during the workshop and to Ditlitabmas Dikti Kemendikbud and Sanata Dharma University for a financial support.

\end{document}